\documentclass[aps,prb,twocolumn,groupaddress,floatfix]{revtex4}

\usepackage{graphicx}
\usepackage{dcolumn}
\usepackage{bm}
\usepackage{amsmath}
\usepackage{multirow}

\newcolumntype{f}[1]{D{.}{.}{#1}}

\begin{document}

\title{X-ray absorption Debye-Waller factors from {\it ab initio}
molecular dynamics}
\date{\today}
\author{F. D. Vila}
\affiliation{Department of Physics, University of Washington, Seattle, WA 98195}
\author{V. E. Lindahl}
\affiliation{Department of Physics, University of Washington, Seattle, WA 98195}
\author{J. J. Rehr}
\affiliation{Department of Physics, University of Washington, Seattle, WA 98195}

\begin{abstract} 

 An {\it ab initio} equation of motion method is
introduced to calculate the temperature-dependent
mean square vibrational amplitudes which appear in
the Debye-Waller factors in x-ray absorption, x-ray scattering, and
related
spectra.  The approach avoids explicit calculations of phonon-modes,
and is based instead on calculations of the displacement-displacement
time correlation function from {\it ab initio} density functional theory
molecular dynamics simulations. The method also yields the
vibrational density of states and thermal quantities such as the
lattice free energy.  Illustrations of the method are presented for
a number of systems and compared with other methods and
experiment.

\end{abstract}

\date{\today}

\maketitle

\section{Introduction}
\label{sec:intro}

Thermal vibrations give rise to exponentially damped
Debye-Waller (DW) factors $\exp[-W(T)]$ in x-ray absorption spectra (XAS),
x-ray diffraction (XRD), and related spectra.
For example, in x-ray absorption fine structure spectra (XAFS)
$W(T) \approx 2 k^2 \sigma_R^2(T)$ where
$\sigma_R^2=\langle[({\vec u}_R -{\vec u}_0)\cdot {\hat{R}}]^2\rangle$
refers to the mean square relative displacement (MSRD)
of a given bond ${\vec R}$,
$k$ is the photo-electron wave number,
and $T$ the absolute temperature.  \cite{crozier88}
In XRD, and similarly in neutron diffraction (ND) and the M\"ossbauer effect, 
$\sigma^2(T) =\langle ({\vec u}\cdot { \hat k}{})^2\rangle$
is the mean-square displacement of an atom along the momentum
transfer vector $\vec k$,
$ {\vec u}$ being the instantaneous displacement vector.
%In contrast, in XAFS, which is the primary focus of this work,
Due to their strong  variation with
temperature, energy, and the geometrical structure of a material,
accurate DW factors are crucial to a quantitative
analysis of XAS; conversely, the lack of precise Debye-Waller
factors is one of the main limitations to accurate structure determinations
from experiment, especially for coordination numbers.

Various methods have been developed for obtaining these DW factors.
Phenomenological models, e.g.,  correlated Einstein and
Debye models \cite{sevillano79,hung97} are widely used in fitting
but are often only semi-quantitative.  More generally, they can
be calculated in terms of Debye-integrals over
appropriate projected vibrational densities of states (VDOS).
In small molecules, explicit sums over modes can
be used to calculate the VDOS.\cite{Loeffen1996403,Dimakis2006L87}
Such sums can also be used for periodic solids, both for
crystallographic Debye-Waller factors and other thermodynamic
quantities.\cite{baroni01,lee95,gonze96} For complex materials, however,
calculating and summing over modes can  be a computational bottleneck.
As an alternative, a Lanczos algorithm can be used to evaluate the VDOS,
starting from a dynamical matrix (or Hessian), that can be obtained either
from force-field models,\cite{poiarkova01,krappe02} or first principles DFT
calculations.\cite{vila07} At high temperatures, brute force classical
MD (or DFT/MD) methods can also be used to obtain moments of vibrational
distribution functions,\cite{vila08} but such methods can fail at
low temperatures when quantum statistics dominate. First principles
DFT methods can also be computationally demanding, especially in complex
systems.

In an effort to speed up the calculations, we present here a first
principles approach based on an {\it ab initio} equation of motion (AEM)
approach using DFT molecular dynamics calculations of 
displacement-displacement time-correlation functions.
The method is a generalization of the equation of motion 
method \cite{rehralben} for calculations of the VDOS,
which was adapted for calculations of Debye-Waller
factors based on  force-field models.\cite{poiarkova99} However
since accurate force-field models are not generally available, especially
for complex molecules and solids, DFT or other {\it ab initio}
methods are needed.
Because $\sigma_R^2$ depends primarily on the vibrational structure in
the local environment around a given bond $R$, the calculations can
be carried out using relatively small clusters of atoms, without the use
of periodic boundary conditions or other symmetry considerations.
Thus the approach is applicable to general aperiodic materials.

\section{Equation-of-motion method}
\subsection{Formalism}
%\label{sec:meth}

The theory used in the present study is a first principles extension of the
equation of motion approach \cite{rehralben,poiarkova99} for calculations of the
VDOS and thermodynamic quantities that can be expressed as Debye-integrals over
the VDOS. Our {\it ab initio} equation of motion (AEM) extension builds in
dynamical structure in terms of first principles DFT calculations for a general
structure, but does not rely on explicit calculations of the dynamical matrix (DM). 
The technique builds in Bose-Einstein statistics, and allows one to
calculate the DW factors and related thermal properties either in real-time or
the frequency domain. The AEM method has a number of computational advantages.
It can be efficient even for large systems, since the method is local and
diagonalization of huge matrices is not necessary. Also the computational time
scales linearly with the size of a cluster. Anharmonic effects such as
lattice expansion can be added using a cumulant expansion.\cite{vila07}

Our AEM method is based on calculations of the displacement-displacement
correlation function in real time, using solutions of the $3N$ coupled
Newton's equations of motion with 
DFT/MD methods. Such correlation functions are Fourier transforms
of projected vibrational densities of states (VDOS), which are defined
uniquely by the initial conditions. Physically the VDOS  can be interpreted
as the ``sound" of a lattice ``plucked" along a given
set of initial displacements.
Here $N$ is the number of atoms in the system which is  centered
within the region of interest and typically a few near-neighbors in
radius.  %Within  the DFT/MD approach
Regarding the total lattice
potential energy $\Phi$ of the crystal lattice as a function of the
local atomic displacements $\vec u_i$ from  their thermal-equilibrium
positions $\vec R_i$, and making use of a quasi-harmonic approximation,
%at a given temperature $T$,
the equations of motion can be written as
\begin{equation}
\label{em}
{\frac{ d^2  Q_{i\alpha}(t)}{ dt^2}} = 
- \sum_{k\beta} D_{i\alpha,k\beta} Q_{k\beta},
\end{equation}
with given initial displacements $\vec u_i(0),\ (i=1\dots N)$,
and zero initial velocities $\dot{\vec u}_i(0)=0$.
%in real-time using the Velocity-Verlet
%algorithm.\cite{verlet67a,verlet67b}
Here ${\vec Q_i} = {\vec u_i} \sqrt{M_i}$ denote reduced displacements
at site $i$ where $M_i$ is the atomic mass,
and $D_{i\alpha,k\beta}={\Phi_{i\alpha,k\beta}/ \sqrt{M_i M_k}}$ is
the dynamical matrix
of order $3N \times 3N$. The matrix $\Phi_{i\alpha,k\beta}$ consists of
second derivatives of the potential energy with respect to the atomic
displacements $u_{i\alpha}$ and $u_{k\beta}$, where $i$, $k$ are atomic sites and $\alpha, \beta = \left( x, y, z \right)$.
% taken at the thermal-equilibrium
%configuration, which is appropriate for the quasi-harmonic approximation.
Formally, the reduced displacement vectors ${\vec Q}_{i}$ can be
%Upon substituting the canonical
expanded in normal coordinates $q_\lambda$ and eigenmodes $\lambda$ as
\begin{equation}
\label{qs}
{\vec Q_i} = \sum_\lambda{\vec \epsilon_i}(\lambda)\, q_\lambda .
\end{equation}
Substituting this relation into Eq.\ (\ref{em}),
%\noindent into the definition of the mean square fluctuation in the effective 
%path length $R_j$,
leads to a standard eigenvalue problem for the normal modes,
\begin{equation}
\label{nmodes}
\omega _\lambda ^2\, \epsilon_{i\alpha} (\lambda)= 
\sum_{k\beta} D_{i\alpha,k\beta}\, \epsilon_{k\beta}(\lambda) .
\end{equation}
After evaluating the thermal averages using Bose-Einstein statistics,
one obtains for the normal coordinates
\begin{equation}
\label{bes}
\omega_\lambda^2 \langle q_\lambda \rangle ^2 =
\langle n(\omega_\lambda) + \frac{1}{2} \rangle \hbar \omega_\lambda
= \frac{\hbar \omega_\lambda}{ 2} \coth \frac{\hbar \omega_\lambda \beta}{
2}.
\end{equation}
\noindent
%and generalizations to non-equilibrium occupation-numbers is straightforward.

 In applying these results for calculations of interest here, it is
convenient to define a normalized displacement state 
\begin{equation}
|Q(t)\rangle = |\vec Q_1(t),\vec Q_2(t), \dots \vec Q_N(t) \rangle.
\end{equation}
For example, for the MSRD for a given near-neighbor bond
$(\vec{0}, \vec{R})$,\cite{poiarkova99}
the initial displacement state $\vert Q_R(0) \rangle$
has $\vec Q_0(0) = -\sqrt{\mu_R / M_0}\hat R$;
$\vec Q_R(0) = +\sqrt{\mu_R / M_R}\hat R$,
and otherwise $\vec Q_i=0$,
where $\mu_R = (1/M_R +1/M_0)^{-1}$ is the reduced mass.
%As discussed in Ref. \onlinecite{vila07}, the projected VDOS needed
%for other physical quantities can be obtained by varying the initial
%displacements, as described below.
A frequency domain expression for the MSRD can then  be obtained
from Eqs.\ (\ref{qs}-\ref{bes}) and summing over all modes, i.e.,
\begin{eqnarray}
\label{sigj}
\sigma_R^2&=&\langle[({\vec u}_R -{\vec u}_0)\cdot {\hat{R}}]^2\rangle
\nonumber \\
&=& \frac{\hbar}{2\mu_R} \sum_{\lambda}
\frac{|\langle \lambda|Q_R(0)\rangle|^2}{\omega_\lambda}
\coth {\frac{\beta\hbar\omega_\lambda}{2}} \\
&=& 
\label{sigj_2}
\frac {\hbar}{2 \mu_R} \int_0^{\omega_{max}}\, \frac {d\omega}{ \omega}
             \rho_R(\omega) \coth\frac{\beta\hbar\omega}{2}.
\end{eqnarray}
%On summing over all modes, one obtains a frequency-domain formula for
Here
\begin{equation}
\label{rho_R}
\rho_R(\omega) \equiv \sum_\lambda
\vert \langle \lambda \vert Q_R(0) \rangle \vert ^2
\delta(\omega - \omega_\lambda)
\end{equation}
is the projected VDOS contributing to relative vibrational
motion along $\hat R$ and  $\beta=1/k_BT$.
%$\mu_R$ is an effective reduced mass
%for scattering path $j$
%that insures normalizization and
The maximum frequency  $\omega_{max}$ in Eq.~(\ref{sigj_2}) can be estimated
from the relation
$\omega_{max}\gtrsim z \sqrt{k/\mu_R}$  where
%maximum frequency of the lattice motion,
$z$ is the coordination number and
$k$ is the near-neighbor force constant.
%of the scattering center 
%and its first neighbor.

 In order to obtain an equivalent time-domain expression for the VDOS,
we calculate the cosine-transform of the
displacement-displacement time-correlation function 
$\langle Q_R(t) | Q_R(0)\rangle$
with an {\it ad hoc} exponential damping factor 
that limits the maximum time $t_{max}$ of the integration and
MD runs,
\begin{eqnarray}
\label{rhoR}
\rho_R(\omega) &=& \frac{2}{\pi}\int_0^{t_{max}} \langle Q_R(t) | Q_R(0)\rangle
\cos\omega t\, e^{-\varepsilon t^2} dt \\
%\rho_R(\omega)
\label{rhoR_2}
&\equiv&\sum_\lambda
\vert \langle \lambda \vert Q_R(0) \rangle \vert ^2
\delta_\Delta (\omega - \omega_\lambda).
\end{eqnarray}
Thus as a consequence of  the damping factor,
the projected VDOS of Eq.~(\ref{rhoR_2}) is broadened by
%$\delta_\Delta$ is a
narrow $\delta$-like functions of width $\Delta$
typically chosen to be about 5\% of the bandwidth. This broadening
also smooths the otherwise discrete spectrum of the finite system
used, but has practically no effect on integrated quantities.
%which is introduced as a convergence parameter.
The spectral width $\Delta$ is determined by the
cutoff parameters $\varepsilon= {3 / t_{max}^2}$
and $t_{max}={\sqrt{6} /(\omega_{max} \Delta})$.
These cutoff parameters also focus on the local
environment by cutting off long distance behavior.
The time-correlation function in Eq.~(\ref{rhoR}) is
\begin{equation}
\langle Q_R(t) | Q_R(0)\rangle=
\sum_{i,\alpha}^{n_R} Q_{i\alpha}(t)Q_{i\alpha}(0).
\end{equation}
where $n_R$ is the number of non-vanishing displacements in $|Q_R\rangle$.
Instead of using the 2nd order differential equations in Eq.~(\ref{em}),
in our approach the
displacement state vector $|Q_R(t)\rangle$ is determined by
integrating the equations of motion numerically using
Velocity-Verlet\cite{verlet67a,verlet67b} molecular
dynamics with initial conditions as in $|Q_R(0)\rangle$,
\begin{eqnarray}
\vec R_i(t+\Delta t) &=& \vec R_i(t) + \vec v_i(t) \Delta t +
   \frac{1}{2}~ \vec a_i(t) {\Delta t}^2 ,  \\
\vec v_i(t+\Delta t) &=& \vec v_i(t) +
   \frac{1}{2}~ \left[ \vec a_i(t) + \vec a_i(t + \Delta t) \right] \Delta t ,
\end{eqnarray}
where $\vec R_i(t)$, $\vec v_i(t)$ and $\vec a_i(t)$ are, respectively, the
instantaneous position, velocity and acceleration of atom $i$. The acceleration
$\vec a_i(t) = \vec f_i(t)/M_i$ and $\vec f_i(t)$ is the force on atom $i$.
%At the electronic ground state,
The Hellmann-Feynman theorem ensures that the
forces can be calculated as the expectation value of the analytical
derivative of the Hamiltonian with respect to the nuclear positions.
%Thus, the forces can be calculated analytically and efficiently at each time
%step.
This algorithm is efficient since an explicit calculation of the dynamical
matrix at each time-step is not necessary.
%displacements $|Q_R(0)\rangle$.
%The specific form of the
%initial displacements depends on the multiple-scattering path,
%as defined below. 

Finally a real time expression for the MSRD can be obtained
by substituting Eq.~(\ref{rhoR}) for $\rho_R(\omega)$ into (\ref{sigj_2}) and 
evaluating the Fourier transform, yielding
\begin{eqnarray}
\label{sigjt}
\sigma_R^2(T) &=& 
\frac{\hbar}{ \mu_R\pi} \int_0^{t_{max}} dt 
{\langle Q_R(t) | Q_R(0)\rangle} \nonumber \\
&\times& \ln \left[ (2\sinh\frac{\pi t}{\beta\hbar})^{-1} \right]
e^{-\varepsilon t^2}.
\end{eqnarray}

Eqs.\ (\ref{sigj_2}), (\ref{rhoR}) and (\ref{sigjt}) are the key formulas used
in our AEM
calculations. Throughout this work results obtained with Eqs. (\ref{sigj_2})
and (\ref{rhoR}) will be labeled AEM-FT,
while those obtained with Eq. (\ref{sigjt}) will be labeled AEM-RT.
The form of Eq.~(\ref{sigjt}) shows that, it is not
essential to determine the VDOS  $\rho_R(\omega)$ as an intermediate step, and
hence that $\sigma_R^2(T)$ can be calculated directly from the
corresponding displacement-displacement autocorrelation function.
Note that in the time domain the analog of the Bose-Einstein weight factor
is $-\ln \left[ 2\sinh {\pi t}/{\beta\hbar} \right] = {\pi t}/{\beta\hbar}
- \ln \left[ \exp \left( {2\pi t}/{\beta\hbar} \right) - 1 \right]$.
%$ \ln [1/2\sinh{(\pi t/ \beta\hbar)}] = \ln (1/1 - \pi t/\beta\hbar) -(\pi %t/\beta\hbar)$.
At long times when $\hbar /2\pi t < k_B T$ the weight factor is negative
and reduces to 
$-\pi t/\beta\hbar$ at high temperatures and to $ - \ln{(2\pi t/\beta\hbar)}$ 
at low. Due to the exponential damping, the net time integration limit
$t_{max}$ is usually several
vibrational cycles and typically requires about 25--35 time-steps per cycle
for accuracy to a few percent.
In addition, the singular behavior of the integrands in Eq.\ (\ref{rhoR}) and
(\ref{sigjt}) must be handled with care.
This is especially important at low temperatures due to zero-point motion.
%and the limited time-integration range
%introduces aperiodic trends that
%In the
%time-domain we employ an analogous approach to remove spurious behavior,
%to remove the trends at long times,
Thus in the time-domain, we further stabilize the long time behavior by
convolving the time-correlation function with the inverse Fourier transform of a
smoothed, low-frequency cutoff function
$\Theta(\omega-\omega_c)$, where $\omega_c$ is an appropriate cutoff frequency.
In the frequency domain in Eq.\ (\ref{sigj_2}) we replace the
very low-frequency region with a similar cutoff or a Debye-model
chosen to fit the very low frequency  behavior of $\rho_R(\omega)$.
All the integrals in our implementation of the AEM method are evaluated using 
the trapezoidal rule, which is appropriate for highly oscillatory integrands.

\subsection{ Maximum Entropy Method}
\label{subsec:mem}

Since the MSRDs are obtained from Debye-integrals over the VDOS, a precise
determination of the spectra is not important, as long as the leading moments
are accurate. Thus, as an alternative approach, the projected
density of states can be obtained approximately using the Maximum
Entropy Method (MEM).\cite{numrec} In this approach the VDOS is
approximated as
\begin{equation}
\label{rhomem}
\rho_R(\omega) = {a_0}{ \left| 1 + \sum_{k=1}^{M} a_k e^{i k \omega \Delta
t} \right|^{-1}},
\end{equation}
where $\Delta t$ is the sampling interval in the time domain and $M$ is the
desired order of the approximation. The MEM approach is well suited to
represent phonon densities with sharp resonances, due to the presence of poles
in Eq.~(\ref{rhomem}).
The coefficients $a_k$ can be obtained by solving the system
of linear equations
\begin{equation}
- \sum_{j=1}^{M} \phi_{\left| j-k \right|} a_j = \phi_k \qquad (k=1,\dots M),
\end{equation}
where
\begin{equation}
\phi_l = \frac{1}{N-l} \sum_{i=1}^{N-l}
       \langle Q_R(t_i) | Q_R(0)\rangle \langle Q_R(t_{i+l}) | Q_R(0)\rangle.
\end{equation}
%Here $\langle Q_R(t_i) | Q_R(0)\rangle$ is the autocorrelation function
%sampled at time $t_i$, and
and $N$ is the number of MD evolution steps.
Although the MEM method can be less efficient than the direct
FT approach, we find that it can be more stable in the low frequency region,
since it is less sensitive to
non-periodic trends in the time evolution. The reduced efficiency arises from
the high order of approximation ($M > 200$) needed to achieve an
accurate representation of $\rho_R(\omega)$ at all frequencies. Throughout this work
results obtained with Eqs. (\ref{sigj_2}) and (\ref{rhomem}) will
be labeled AEM-MEM.

\subsection{ Multiple scattering $\sigma_j^2$ }
\label{subsec:ms}

The above real-time AEM method can also be used to calculate
the MSRD $\sigma_j^2$ for a given
XAFS multiple-scattering path $j$ with $n_j$ legs.
%The sum of terms in Eq.~(\ref{defsig}) can be regrouped 
This MSRD corresponds to the mean-square fluctuation in the effective
MS path length $\delta R$ \cite{poiarkova99}
\begin{equation}
\label{sigj2}
\sigma_j^2 \equiv \langle \left( \delta R \right)^2\rangle = 
\left\langle \left[ \sum_{i=1}^{n_j}
{\vec u_i} \cdot \vec\Delta_i
\right]^2 \right\rangle.
\end{equation}
Here $\vec\Delta_i = (\hat R_{ii-} + \hat R_{ii+})/2 $,
where $\hat R_{ii\pm}$ represent the directional unit vectors between the site
$i$ and the sites $i-1$ before and $i+1$ after, along the
multiple-scattering path $j$. 
In analogy with the single scattering results,
%Eq.~(\ref{sigj}-\ref{rhoR_2}),
we obtain expressions similar to Eq.~(\ref{sigj_2}) for 
$\sigma_j^2$ and  Eq.~(\ref{rho_R}) for $\rho_j(\omega)$, but with
the weights in mode $\lambda$ given by
\begin{equation}
\label{sigj3}
\vert \langle \lambda \vert Q_j(0) \rangle \vert ^2 =
\left|\sqrt{\frac{\mu_j}{M_i}}
%\left({\frac{\hat R_{ii-} + \hat R_{ii+}}{2}}\right)
\vec\Delta_i
\cdot {\vec \epsilon_i}(\lambda) \right|^2.
\end{equation}
%\noindent The term in square brackets corresponds to the weight
%${ \vert \langle \lambda \vert Q_j(0) \rangle \vert ^2}$  of a
%given mode $\lambda$ in Eq.~(\ref{rhoj}), and can be interpreted
These weights can be interpreted as the normalized probability that
an initial displacement state $\vert Q_j(0) \rangle$, corresponding
to a multiple-scattering path stretch, is in vibrational
mode $|\lambda\rangle$.  Thus the initial displacements in the state
$\vert Q_j(0) \rangle$ are 
%$\vec Q_1 = \sqrt{\mu_j / M_1}(\hat R_{1,n_j-} + \hat R_{1,2})/2$ and
$\vec Q_i = \sqrt{\mu_j / M_i}\,\vec\Delta_i$,
($i=1,\ldots n_j$).
Here the inverse reduced mass is
\begin{equation}
\label{mu}
\frac{ 1}{\mu_j} \equiv \sum_{i=1}^{n_j} 
\frac{1}{ M_i } |\vec\Delta_i|^2
%\left({\frac{ \hat R_{ii-} + \hat R_{ii+}}{ 2}}\right)^2,
\end{equation}
which is defined so that $\langle Q_j(0) \vert Q_j(0)\rangle =1$ and
$\rho_j(\omega)$ is normalized.

\subsection{Other Dynamical Properties}
%\label{sec:oth_prop}

Other dynamical properties can be obtained similarly,
by generalizing the seed-state $|Q_R(0)\rangle$ appropriately.\cite{vila07}
For example, when the seed state is defined as a single-atom displacement,
the resulting correlation
function yields the mean square atomic displacements $u^2(T)$ in
x-ray scattering DW factors. Also, when all symmetry unique
Cartesian atomic displacements are added, one obtains the total VDOS per site
$\rho_T(\omega)$. This  permits calculations of thermodynamic
functions such as the vibrational free energy per site,\cite{vila07}
\begin{equation}
\label{eq:free_energ}
   F(T) =  3 k_B T \int_{0}^{\infty} d\omega\, \ln \left[{
            {2 \sinh \left(\frac{\beta \hbar \omega}{2} \right)} }\right]
\rho_T(\omega) ,
\end{equation}
where $k_B$ is the Boltzmann constant. Finally, if the $|Q_R(0)\rangle$ seed
state is initialized with atomic displacements perpendicular to $\hat R$ instead
of parallel to it, we can generate the mean-square transverse
displacement $\sigma_{\perp}^2(T)$, which
provides a correction to the lattice expansion.\cite{vila07} 
%the perpendicular component of the DW factor.
%and $\rho_T(\omega)$ is the total phonon  density of modes per site.

\subsection{Computational Details}
\label{sec:meth_comp_det}

The micro-canonical (i.e., NVE) ensemble MD simulations for the applications
presented here were done using {\sc vasp}\cite{vasp96} for the crystalline
systems and {\sc siesta} \cite{artacho99,soler02} for the Zn-imidazole complex.
These codes were chosen on the basis of efficiency, although in principle, any
program capable of NVE dynamics can be interfaced with the AEM codes used
in this work. The {\sc vasp} simulations used standard ultrasoft
pseudopotentials, and were optimized for efficiency in MD runs.
The Ge calculations used a 2$\times$2$\times$2 $k$-point grid
with a plane-wave cutoff of 105 eV, while for ZrW$_2$O$_8$ the grid was
4$\times$4$\times$4 and the cutoff was 297 eV.  The {\sc siesta} calculations
used Troullier-Martins norm-conserving pseudopotentials\cite{Troullier-Martins}
and standard double-$\zeta$ basis sets with a single polarization function
(DZP). The confinement-energy shift defining the numerical atomic orbitals was
10 meV. Finally, the Hartree and exchange-correlation potentials were
represented on a real-space grid with a plane-wave-equivalent cutoff of 120 Ry
within a (18.4 \AA)$^{3}$ cell. Both crystalline and molecular simulations used
the PBE functional.\cite{PBE} We have previously shown that the choice of
exchange-correlation functional plays an important role in obtaining accurate
MSRDs for metallic systems.\cite{vila07} However, here we only focus
on non-metallic and molecular systems, for which the PBE functional
yields reasonable accuracy compared to experiment.\cite{vila07} 

\subsection{Efficiency Considerations}
\label{sec:meth_effic}

The efficiency of the AEM method depends on three factors: 1) The number of
individual MSRDs that need to be computed, 2) the minimum and maximum
frequencies that contribute to the VDOS, and 3) the quality of the
\textit{ab initio} MD. First, if a large number of MSRDs is needed, the
computation of the full DM may be preferable since it yields all necessary DW
factors with minimal additional effort. However, in most XAFS analysis only a
handful of local DW factors need to be known accurately
while those for more distant shells can approximated roughly using
correlated Debye or Einstein models. For example, in the
case of the coordination shell around a metallic center in a complex biomolecule
the AEM approach can provide an efficient alternative to the Lanczos DM
approach. Second, if a given MSRD has similar contributions from low and high
frequency modes, the MD must have a short enough time-step to accurately
represent the high frequency (25--35 steps per cycle) and a total run time with
sufficient cycles of the low frequency (4-8 cycles). Third, the AEM approach
can take advantage of efficient implementations of DFT energies and forces such
as those used here, without relying on analytic second derivatives
needed in the Lanczos DM approach or the equations of motion in
Eq.~(\ref{em}).

Of the applications presented here, results for Ge and
Zn$^{+2}$-tetraimidazole can be more efficiently  treated using the
Lanczos DM approach. In the case of Ge this is due to the simplicity of
the unit cell. In the case of Zn$^{+2}$-tetraimidazole, first there
are a relatively small number of modes and second the modes cover
a broad range of frequencies that would require small time-steps and
a long total simulation time to represent accurately.
%Lanczos DM calculations for this case are several times faster
%than the required MD simulations.
On the other hand, the zirconium tungstate
(ZrW$_2$O$_8$) system, illustrates the definite advantage of the AEM
approach for complex systems,
%since the range of
%frequencies is narrower than in Zn$^{+2}$-tetraimidazole,
since only a handful of MSRDs are needed for XAFS, while the unit cell 
contains hundreds of atoms.  Based on our experience 
with the DM Lanczos approach, we estimate that the AEM approach
would be nearly two orders of magnitude faster than
a dynamical matrix calculation.

%applications
\section{Applications}

\subsection{Germanium}
\begin{figure}[t]
\includegraphics[scale=0.35,clip]{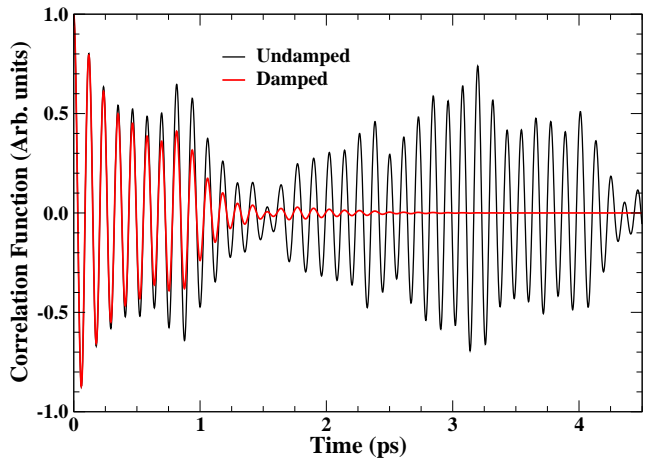}
\caption{\label{fig:ge_corr}
(Color online)
Displacement-displacement correlation function for the
nearest neighbor Ge-Ge bond,
with and without a damping factor $\epsilon=7
\times 10^{-7}\ {\mathrm{fs}^{-2}}$, obtained from a constant
energy molecular dynamics simulation.
}
\end{figure}
As a relatively simple test case, the AEM was applied to a crystalline
germanium system using an 64-atom supercell generated by repeating
2$\times$2$\times$2 times the diamond cubic cell, with the
experimental lattice constant of 5.6575 \AA. The MD simulations
used a 2 fs timestep and a total
simulation time of 4.5 ps. The initial structure was generated by introducing a
4.8\% bond stretch to one of the nearest neighbor pairs in the cell.

%The crystal considered in our study was a 64-atom cluster of Ge of the diamond
%space group with enforced periodic boundary conditions. Calculations were made
%for the single scattering nearest neighbor path at a temperature of 300 K.

\begin{figure}[t]
\includegraphics[scale=0.35,clip]{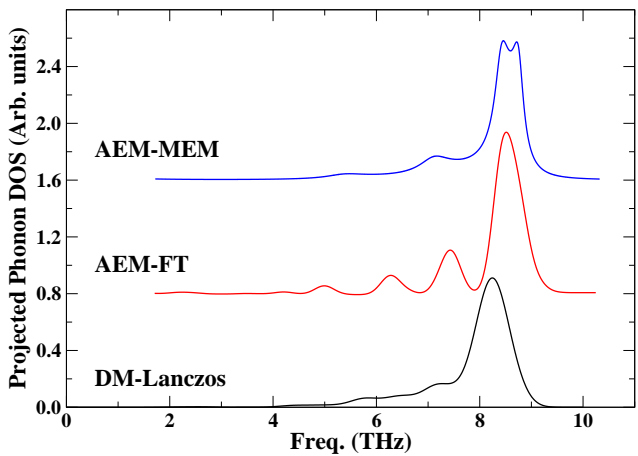}
\caption{\label{fig:ge_rho}
(Color online)
Phonon density of states projected on the 
nearest neighbor Ge-Ge interaction calculated with the
AEM-MEM and AEM-FT approaches, and for comparison, the
broadened Lanczos DM results.
}
\end{figure}
The correlation function resulting from the Velocity-Verlet time evolution
is shown in
Fig.~\ref{fig:ge_corr}. As expected, the oscillations are dominated by a
single mode with a period of about 117 fs, associated with the Ge-Ge
optical mode stretch.
This dominant behavior can also be observed in the VDOS
shown in Fig.~\ref{fig:ge_rho}, where the optical modes are centered
at about 8.5 THz. %  while the
An integration time of about 2 ps is adequate
to obtain phonon-spectra with a spectral broadening of about 5\%.
The centroid of the VDOS is located at about 8 THz,
in good agreement with Einstein models for the nearest neighbor
single-scattering path with an Einstein frequency of 7.55 THz.\cite{dalba99} It
should be noted that although the integration time for optical mode is well
above that needed for convergence, the net integration time 
for lower frequencies around 5 THz is just adequate.  Due to the singular
behavior in Eq.~(\ref{sigj}), an adequate time integration for
the lower frequency components is essential, and is especially important
at low temperatures for some of the systems discussed in the next section. 
%The correlation function was obtained using the density functional theory based
%implementation VASP. The result is illustrated in figure 3. The oscillating
%system was iterated forward 4500 fs in time. From the figure,  one can note the
%presence of a dominating frequency and the period of those oscillations is
%around 117 fs. This leaves us with about 38 full vibrational cycles. In
%practice however, one would probably only include ,10 cycles or less for
%computational efficiency.

\begin{figure}[t]
\includegraphics[scale=0.35,clip]{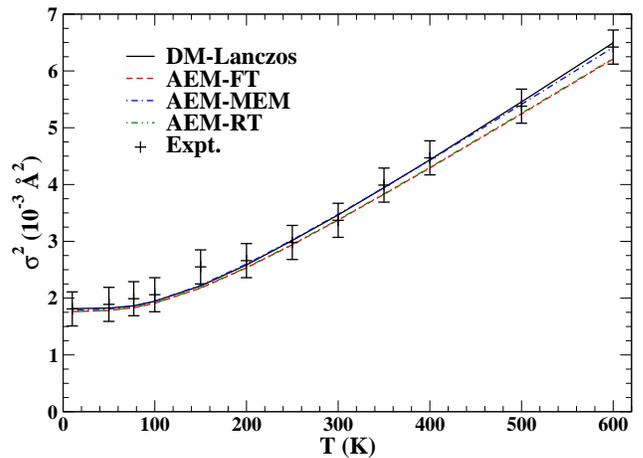}
\caption{\label{fig:ge_s2}
(Color online)
Nearest neighbor Ge-Ge MRSD calculated with the
AEM-FT, AEM-MEM and AEM-RT approaches, and for comparison,
Lanczos DM and experimental\cite{dalba99}
results. The experimental results are shifted as in Ref.
\onlinecite{vila07}.
}
\end{figure}

The MSRDs calculated for the nearest neighbor Ge-Ge bond are shown in Fig.~\ref{fig:ge_s2}.
The agreement with experiment is quite good, with an average
error of 4\% for the AEM-FT approach and 2\% for the AEM-MEM approach. For
comparison, the DM-Lanczos approach has an average error of
2\%. Fig.~\ref{fig:ge_s2} also shows the results obtained with the real-time
approach of Eq.~(\ref{sigjt}) and a frequency cutoff of 1.7 THz as in the FT and
MEM approaches. As expected, given the formal equivalence between
Eq.~(\ref{sigjt})
the FT approach with an intermediate calculation of the VDOS, the results are
nearly identical. 

To explore the accuracy and efficiency of the AEM-FT and AEM-MEM approaches,
we have also integrated the correlation function both for shorter times and 
for larger time steps. For Ge we find that the total integration time
can be reduced to about 1 ps without significant loss of accuracy.
This corresponds to about
10 periods of the 8.5 THz dominant frequency. For integration times of about 500 fs
the mean error for the MSRD increases to 8\% for the FT approach and to 16\% for
MEM. From the point of view of the length of the time step, both the FT and MEM
approaches are extremely resilient. In both cases the mean errors for the Ge
MSRD remain constant with time steps up to 24 fs. This corresponds to
approximately five samples per period of the 8.5 THz frequency. Such large time
steps, however, might not be feasible within the MD simulation itself due to
loss of energy conservation in the Verlet algorithm.

%The important quantity for obtaining the Debye- Waller factors, is determined
%to be 3.23 $\times$ 10$^{-3}$\AA. The experimental value available is 3.50
%$\times$ 10$^{-3}$\AA. Since our accuracy requirements lie in the 10-20\%
%range, this degree of agreement is satisfactory. We can also study the
%temperature dependence of our result. Fig.~5 shows the temperature dependence
%of  determined to be 3.23 $\times$ 10$^{-3}$\AA. The experimental value
%available is 3.50 $\times$ 10$^{-3}$\AA. at temperature T relative to
%$\sigma^2$ at 0 K, and this difference is the quantity that can usually be
%extracted from experiment. The pink line shows results from using the
%correlated Debye model and the blue line the experimental data. It is clear
%that the correlated Debye model performs poorly compared to the AEM method,
%shown in blue. The effect of the time length and time step length of the
%correlation function was also studied. The effect of changing the time length
%from 4500 fs to 700 fs  changed $\sigma^2$ from 3.23 $\times$ 10$^{-3}$\AA. to
%3.60 $\times$ 10$^{-3}$\AA. Doubling the time step length from 2 fs to 4 fs
%changed $\sigma^2$ from 3.23 $\times$ 10$^{-3}$\AA.  to 3.46 $\times$
%10$^{-3}$\AA. while a tripling resulted in 3.66 $\times$ 10$^{-3}$\AA.

%The relative shift in $\sigma^2$ due to time length effects is illustrated in
%Figure 6. Here, the $y$-axis is the relative ``error" in that is introduced due
%to decreased time length. Note that the $x$-axis shows the number of time steps
%and not the actual time.

\begin{figure}[t]
\includegraphics[scale=0.35,clip]{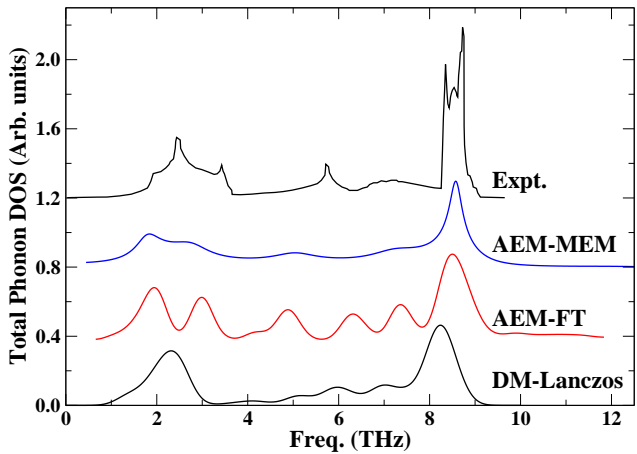}
\caption{\label{fig:ge_rho_tot}
(Color online)
Total phonon density of states for Ge calculated with the
AEM-MEM and AEM-FT approaches, and compared to
results from the Lanczos DM and experiment.\cite{nelin72}
}
\end{figure}

As an example of other dynamical quantities that can be obtained with the AEM
approach, Fig.~\ref{fig:ge_rho_tot} shows the total phonon density of states for
Ge calculated with the FT and MEM approaches. For comparison broadened
dynamical matrix Lanczos and experimental\cite{nelin72} results are also included. This
VDOS was obtained by applying a single atomic displacement along the $\hat{x}$
axis, as described in \ref{subsec:ms}, and by propagating as for 
$\sigma^2$ for 4.5 ps.
%Given that integrals over the VDOS are
%needed for thermal properties, precise details are not as important
%as integrated properties such as the mean and lower order moments.
Overall, the
centroid of the DOS is accurately reproduced by all methods: The centroid of the
experimental DOS is located at 5.8 THz, while the FT and MEM approaches place it
at 6.0 and 5.7 THz, respectively. The spread (i.e., 2nd moment) of the DOS
is also well reproduced
with the FT and MEM, giving 2.8 and 2.9 THz, respectively, versus 2.6 THz in the
experiment. Finally, all methods reproduce the positions and weights of main
features of the experimental VDOS quantitatively. On average the positions of
the peaks deviate by at most 0.4 THz (i.e., about 4\% of full bandwidth)
and the relative weights are within 5\% of
those observed in experiment.

\begin{figure}[t]
\includegraphics[scale=0.35,clip]{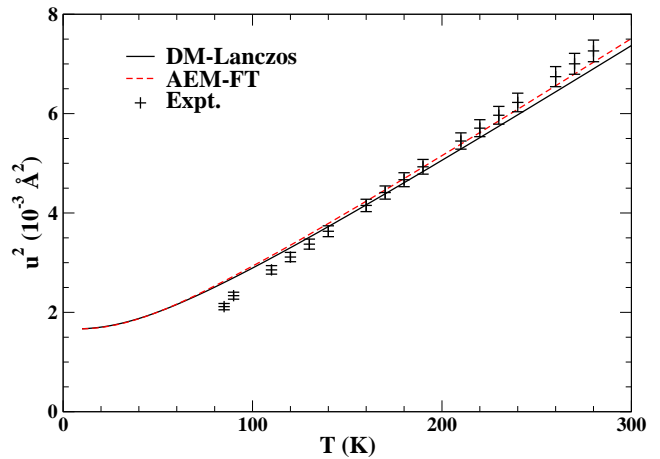}
\caption{\label{fig:ge_u2}
(Color online)
Mean square atomic displacement for Ge calculated with the AEM-FT
approach using a single atomic displacement, and for comparison, Lanczos
DM and experimental\cite{Peng:zh0008} results.
}
\end{figure}

The accuracy of the total VDOS can also be gauged by comparing with the
experimentally measured atomic MSD $u^2$ for Ge.
Fig.~\ref{fig:ge_u2} shows the MSD computed using the total VDOS
shown in Fig.\
\ref{fig:ge_rho_tot}. The AEM results are in excellent agreement with those
obtained with the full DM Lanczos approach and in good agreement
with the available experimental results \cite{Peng:zh0008} except at 
low temperatures.

%The calculated projected and total VDOS is presented in figure 4 in comparison
%with experimental data for the total VDOS6. Just as with the correlation
%function, one can note the presence of a dominating frequency at about 53 T
%rad/s, in all three spectra. Also, the calculated versus experimental total
%VDOS compare reasonably well considering the general features of the plots. The
%larger width of the calculated total VDOS compared to the projected one, is
%most likely a time length effect, since for this calculation the correlation
%function was only determined for about 1200 fs. The more confined the
%correlation function is in time, the more its transform will spread out in
%frequency space.

\subsection{Zn$^{+2}$-tetraimidazole}

\begin{figure}[t]
\includegraphics[scale=0.55,clip]{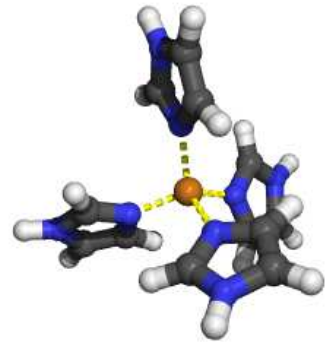}
\caption{\label{fig:znimid}
(Color online)
Structure of Zn$^{+2}$-tetraimidazole.
}
\end{figure}

\begin{figure}[t]
\includegraphics[scale=0.35,clip]{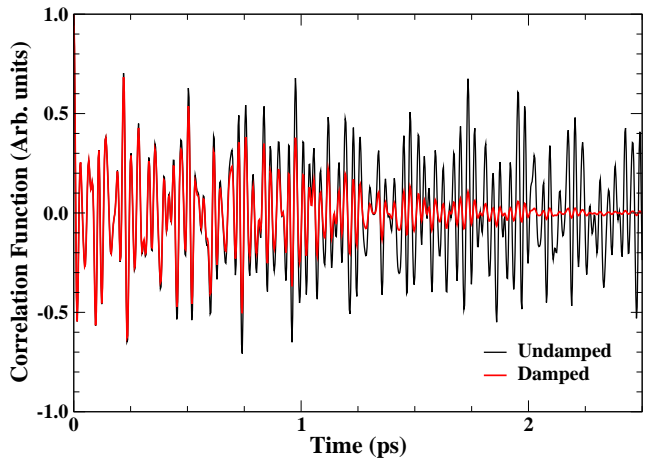}
\caption{\label{fig:znl_corr}
(Color online)
Displacement-displacement correlation function for the
nearest neighbor Zn-N interaction in Zn$^{+2}$-tetraimidazole
with and without a damping factor $\epsilon=6
\times 10^{-7}\ {\mathrm{fs}^{-2}}$, obtained from a constant
energy molecular dynamics simulation.
}

\end{figure}

\begin{figure}[t]
\includegraphics[scale=0.35,clip]{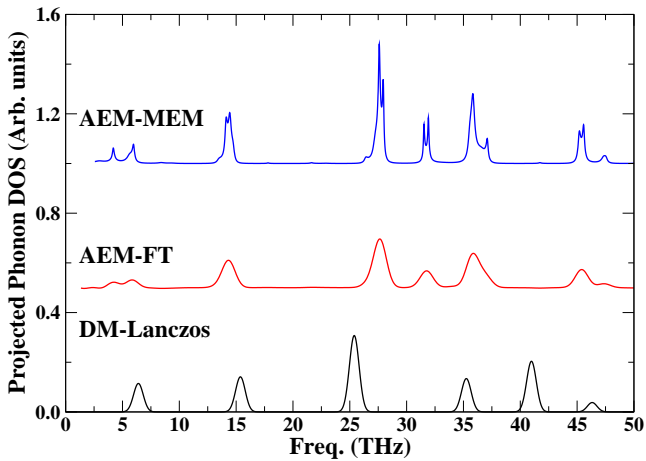}
\caption{\label{fig:znl_rho}
(Color online)
Phonon density of states projected on the 
nearest neighbor Zn-N interaction
in Zn$^{+2}$-tetraimidazole calculated with the
AEM-MEM and AEM-FT approaches, and for comparison, the
broadened Lanczos DM results.
}
\end{figure}

\begin{figure}[t]
\includegraphics[scale=0.35,clip]{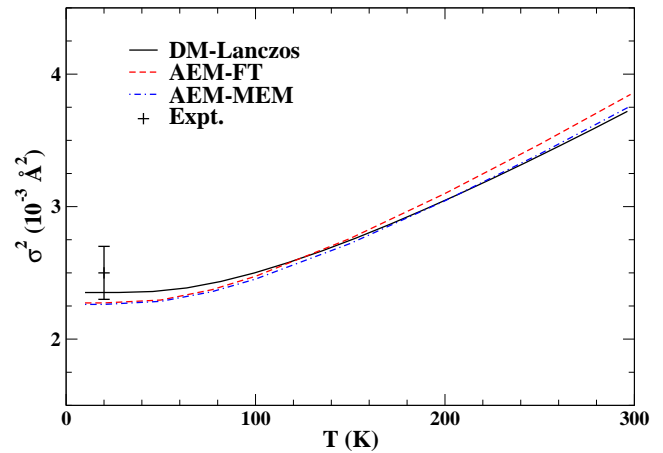}
\caption{\label{fig:znl_s2}
(Color online)
Nearest neighbor Zn-N MRSD in Zn$^{+2}$-tetraimidazole
 calculated with the
AEM-MEM and AEM-FT approaches, and for comparison,
Lanczos DM and experimental\cite{Loeffen1996403}
results.
}
\end{figure}

As an example of a complex molecule,
Zn$^{+2}$-tetraimidazole was simulated using the full structure shown in
Fig.~\ref{fig:znimid}.
This structure was optimized in {\sc siesta} and one of the equivalent Zn-N
bonds was distorted with a 3.4\% bond stretch. The MD simulations used a 3 fs
timestep and a total simulation time of 3.9 ps. Given its large number of
degrees of freedom, the dynamics of Zn$^{+2}$-tetraimidazole are significantly
more complicated than those of Ge. This can be seen in the correlation function
shown in Fig.~\ref{fig:znl_corr}, which exhibits a superposition of several
modes.
The dominant contributions can be analyzed by examining the
VDOS in Fig.~\ref{fig:znl_rho}.
The DM approach exhibits three dominant frequencies at 5, 13 and 25 THz, which
contribute 32, 18 and 24\%, respectively, of the MRSD value. It is interesting
to note that the weight of the associated poles is 9, 13 and 31\%, further
highlighting the importance of the correct representation of the low frequency
modes. In principle, the Zn-N path should be dominated by low frequency
Zn-ligand tetrahedral modes. Loeffen \textit{et al.}\cite{Loeffen1996403} find
that these modes appear at about 6.5 THz, in fair agreement with our principal
contribution at 5 THz. Although the VDOS calculated with the AEM-FT and
AEM-MEM approaches are in good agreement with each other, they have small
differences with respect to the Lanczos DM VDOS.
For instance, the mode at 25 THz is blueshifted about 2 THz in the
real-time approaches.
Fig.~\ref{fig:znl_s2} shows that the agreement between the MSRDs calculated
from the different VDOS is quite good. At 8\% error, the theoretical results are
less accurate than those obtained for Ge. They are, however, still within the
error margins of the available experimental value at 20K. The larger error is
likely due to the quality of the basis set used in the {\sc siesta}
calculations.

\begin{figure}[ht]
\includegraphics[scale=0.30,clip]{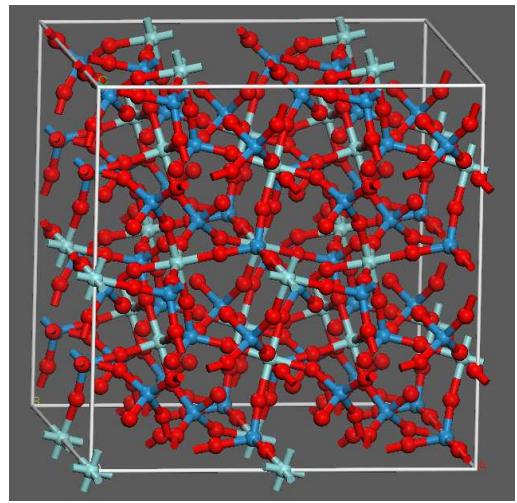}
\caption{\label{fig:zrw2o8}
(Color online)
Structure of 2$\times$2$\times$2 supercell of zirconium tungstate
ZrW$_2$O$_8$ (Zr: light blue, W: dark blue, O: red). 
}
\end{figure}
\begin{figure}[ht]
\includegraphics[scale=0.35,clip]{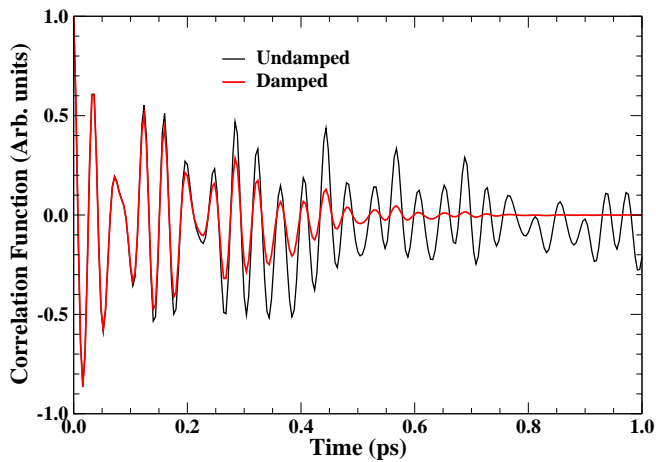}
\caption{\label{fig:zr_corr}
(Color online)
Displacement-displacement correlation function for the
short Zr-O bond nearest neighbor interaction,
in ZrW$_2$O$_8$
with and without a damping factor $\epsilon=6
\times 10^{-6}\ {\mathrm{fs}^{-2}}$, obtained from a constant
energy molecular dynamics simulation.
}
\end{figure}

\subsection{ZrW$_2$O$_8$}

Our final example is zirconium tungstate (ZrW$_2$O$_8$), a ceramic that exhibits
negative thermal expansion (NTE). This system is quite challenging, having a
complex unit cell that puts the calculation of the DM for the Lanczos
approach beyond the reach of our current implementation and computational 
capabilities. Here we have applied the AEM approach
to a 352-atom supercell (Fig.\ \ref{fig:zrw2o8}) made of 2$\times$2$\times$2
repetitions of the unit cell.
The simulations used the experimental unit cell lattice constant of 9.1546 \AA\
and a timestep of 4 fs, for a total simulation time of 1.5 ps. ZrW$_2$O$_8$ has
several interactions of interest, including Zr-Zr, W-W, W-O and two
inequivalent nearest-neighbor Zr-O bonds with distances 2.03 and 2.11 \AA. In
principle, any of these interactions can be studied using the AEM approach. As a
proof of principle here we study the $\sigma^2$ of the shortest of the Zr-O
bonds by using an initial structure corresponding to a 3.8\% bond stretch. 

\begin{figure}[t]
\includegraphics[scale=0.35,clip]{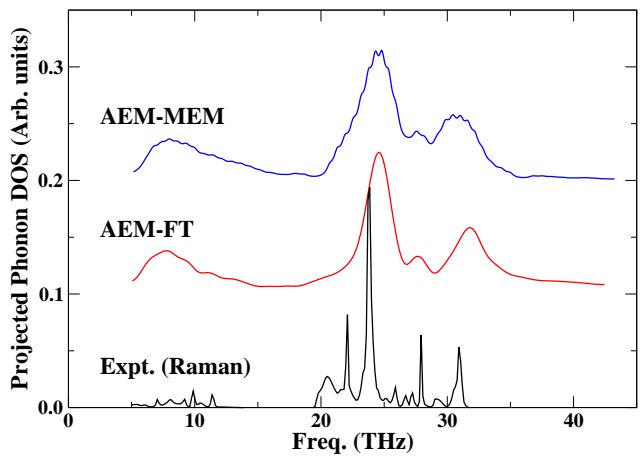}
\caption{\label{fig:zr_rho}
(Color online)
Phonon density of states projected on the 
short Zr-O bond nearest neighbor interaction of ZrW$_2$O$_8$,
calculated with the
AEM-FT and AEM-MEM approaches, and for comparison the experimental
Raman spectrum.\cite{PhysRevB.67.064301} 
}
\end{figure}

For a Zr-O distortion, the dynamics of ZrW$_2$O$_8$  are not as complex as those
observed for Zn$^{+2}$-tetraimidazole. The correlation function (Fig.\
\ref{fig:zr_corr}) is mostly dominated by a mode with a 40 fs period 
superposed on a mode with a period approximately three times longer. Visual
inspection of the MD trajectory reveals that the 40 fs mode is
associated principally with the longitudinal Zr-O stretch mode. 
These vibrational modes can be clearly seen at about 25 and 8 THz,
respectively, in the VDOS shown in Fig.\ \ref{fig:zr_rho}. As in the
previous examples, the agreement between the AEM-MEM and AEM-FT approaches
is very good. The agreement with the mode frequencies observed in the experimental Raman
spectrum is also quite good. The FT and MEM VDOS show modes at approximately
7.7, 24.5, 27.7 and 31.7 THz, compared to the experimental peaks at 5.7-11.8,
23.8, 27.9 and 31.0 THz. The 5.7-11.8 THz peaks are associated mostly with modes
located on the tungstate ion and with some low frequency WO$_4$
modes.\cite{PhysRevB.67.064301} The 23.8, 27.9 and 31.0 THz peaks correspond
exclusively to asymmetric WO$_4$ modes. It is interesting to note that the
dynamics of this system are quite complex. Since the WO$_4$ units are very
stiff, the ZrO$_6$ units must rotate as the WO$_4$ units translate.
\cite{PhysRevB.68.014303} Thus, a simple distortion of the Zr-O bond is able
to activate both the low and high frequency modes.

\begin{figure}[ht]
\includegraphics[scale=0.35,clip]{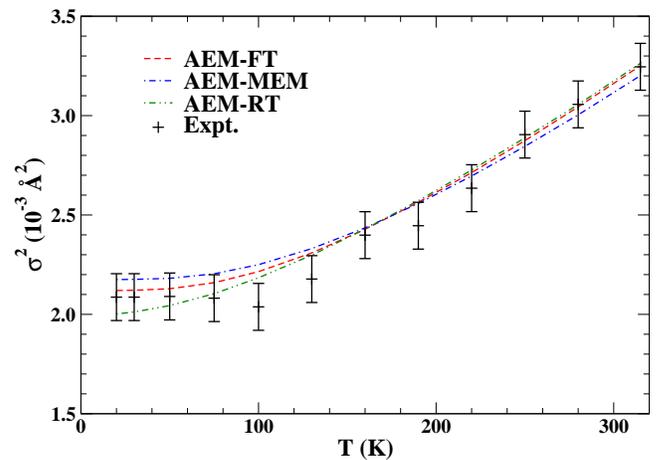}
\caption{\label{fig:zr_s2}
(Color online)
MRSD of the shortest nearest neighbor Zr-O
bond in ZrW$_2$O$_8$ calculated with the
AEM-FT, AEM-MEM and AEM-RT approaches, and for comparison experimental
\cite{PhysRevB.68.014303} results.
}
\end{figure}
Fig. \ref{fig:zr_s2} shows the nearest-neighbor Zr-O MSRD as a function of
temperature. As expected, given the similarity of their VDOS,
the FT and MEM values are in very good agreement. The direct integration
RT approach also agrees well with the FT approach, at least for higher
temperatures. The agreement with experiment\cite{PhysRevB.68.014303} is also
quite good, with all theories falling within the experimental error bars for
most of the temperature range. The largest disagreement occurs in the 80-140K
region where other MSRDs (W-W and Zr-Zr) are known to have an anomaly that is
likely related to the NTE.\cite{PhysRevB.68.014303}

\section{Conclusions}
\label{sec:concl}

We have introduced an {\it ab initio} equation of motion (AEM) method  for
calculations of the MSRDs $\sigma^2$, needed for  Debye-Waller factors in
x-ray absorption, x-ray scattering, and related spectra. The method is based on
calculations of  displacement-displacement time correlation functions from {\it
ab initio} density functional theory molecular dynamics simulations, using the
Velocity-Verlet time-evolution algorithm. Thus the approach avoids the need for
explicit calculations of phonon-modes or the dynamical matrix. The AEM method
builds in Bose-Einstein statistics and yields the vibrational density of states
(VDOS) as either cosine Fourier Transforms of displacement-displacement
correlation functions or
through the Maximum Entropy Method. The MSRDs and other thermal quantities such
as the lattice free energy, are obtained in terms of Debye-integrals over the
VDOS. Alternatively, the MSRDs can be computed directly from the correlation
functions by using the time-domain counterpart of the Bose-Einstein weight
factor. Application of the method to a number of systems show that the approach
is computationally advantageous for large, complex systems, and is in
quantitative agreement with other methods and with experimental results.

\begin{acknowledgments}
We thank J. Kas, J. Vinson, S. Williams, and F. Bridges for comments
and suggestions.
This work is supported in part by NSF Grant PHY-0835543 (FDV and JJR).
One of us (VEL) thanks the REU program at the University of Washington
in summer 2010, which is supported by NSF REU Grant PHY-0754333, where
part of this work was carried out.
\end{acknowledgments}

%\bibliography{dweom}
%\bibliographystyle{apsrev}

\end{document}